\documentclass[twoside]{article}
\usepackage{qic,epsfig}

\usepackage{cite}
\usepackage{booktabs}

\usepackage{multirow}
\usepackage[cmex10]{amsmath}
\usepackage{braket}
\usepackage{bm}
\usepackage{graphicx}
\usepackage{rotating}
\usepackage{enumerate}
\usepackage{subfigure}
\usepackage[pdftex]{color}
\newcommand{\ignore}[1]{}

\input{Qcircuit.tex} % plot quantum circuits

\begin{document}
\setlength{\textheight}{8.0truein}    %FOR 2ND PAGE ONWARDS

\runninghead{Performance and Error Analysis of Knill's Postselection Scheme in a Two-Dimensional Architecture}
            {C.-Y. Lai, G. Paz, M. Suchara, and T.A. Brun}

\normalsize\textlineskip
\thispagestyle{empty}
\setcounter{page}{1}

%\copyrightheading{Vol.}{No.}{Year}{Page Nos.}
%\copyrightheading{0}{0}{2003}{000--000}

\vspace*{0.88truein}

\alphfootnote

\fpage{1}
\centerline{\bf
%%%%%%%%%%%%%%%%%%%%%
%Put in titiles here
%%%%%%%%%%%%%%%%%%%%%
PERFORMANCE AND ERROR ANALYSIS OF KNILL'S }
\vspace*{0.035truein}
\centerline{\bf POSTSELECTION SCHEME IN  A TWO-DIMENSIONAL ARCHITECTURE}
\vspace*{0.37truein}
\centerline{\footnotesize
%%%%%%%%%%%%%%%%%%%%%%%%%%%%%%%%%%%%
%put authors' name and address here
%%%%%%%%%%%%%%%%%%%%%%%%%%%%%%%%%%%%
Ching-Yi Lai}
\vspace*{0.015truein}
\centerline{\footnotesize\it Department of Electrical Engineering, University of Southern California}
\baselineskip=10pt
\centerline{\footnotesize\it Los Angeles, CA 90089 U.S.A.}
\vspace*{10pt}
\centerline{\footnotesize Gerardo Paz}
\vspace*{0.015truein}
\centerline{\footnotesize\it Department of Physics and Astronomy, University of Southern California }
\baselineskip=10pt
\centerline{\footnotesize\it Los Angeles, CA 90089 U.S.A.}
\vspace*{10pt}
\centerline{\footnotesize Martin Suchara}
\vspace*{0.015truein}
\centerline{\footnotesize\it Computer Science Division, University of California at Berkeley }
\baselineskip=10pt
\centerline{\footnotesize\it Berkeley, CA 94720 U.S.A.}
\vspace*{10pt}
\centerline{\footnotesize Todd A. Brun}
\vspace*{0.015truein}
\centerline{\footnotesize\it Department of Electrical Engineering, University of Southern California }
\baselineskip=10pt
\centerline{\footnotesize\it Los Angeles, CA 90089 U.S.A.}
\vspace*{0.225truein}
%\publisher{(received date)}{(revised date)}

\vspace*{0.21truein}
\abstracts{
%%%%%%%%%%%%%%%%%%%%
% put abstract here
%%%%%%%%%%%%%%%%%%%%
Knill demonstrated a fault-tolerant quantum computation scheme based on concatenated error-detecting codes and postselection with a simulated error threshold of $3\%$
over the depolarizing channel.
%We design a two-dimensional architecture for fault-tolerant quantum computation based on Knill's postselection scheme.
We show how to use
Knill's postselection scheme in a practical two-dimensional quantum
architecture that we designed with the goal to optimize the error correction properties, while satisfying important architectural constraints.
In our 2D architecture, one logical qubit is embedded in a tile consisting of $5\times 5$ physical qubits.
The movement of these qubits is modeled as noisy SWAP gates and the only physical operations that are allowed are local one- and two-qubit gates.
We evaluate the practical properties of our design, such as its error threshold, and compare it to the concatenated Bacon-Shor code and the concatenated Steane code.
Assuming that all gates have the same error rates, we obtain a threshold of $3.06\times 10^{-4}$ in a local adversarial stochastic noise model, which is the highest known error threshold for concatenated codes in 2D.
We also present a Monte Carlo simulation of the 2D architecture with depolarizing noise and we calculate a pseudo-threshold of about $0.1\%$.
With memory error rates one-tenth of the worst gate error rates, the threshold for the adversarial noise model, and the pseudo-threshold over depolarizing noise, are $4.06\times 10^{-4}$ and $0.2\%$, respectively.
In a hypothetical technology where memory error rates are negligible, these thresholds can be further increased by shrinking the tiles into a $4\times 4$ layout.}{}{}

\vspace*{10pt}
\keywords{fault-tolerant quantum computation, quantum error correction}
\vspace*{3pt}
%\communicate{to be filled by the Editorial}

\vspace*{1pt}\textlineskip	%) USE THIS MEASUREMENT WHEN THERE IS
\section{Introduction}	        %) A SECTION HEADING
\vspace*{-0.5pt}
\noindent
%%%%%%%%%%%%%%%%%%%%%%%%%%%%%%%%
%put the text of the paper here
%%%%%%%%%%%%%%%%%%%%%%%%%%%%%%%%
\label{Introduction}

Quantum error correction~\cite{Shor96,AB97:FQC:258533.258579,DS96,KLZ96,Got97,Got97a,Ste97L,Ste97a,Pre98c,Ste99N} is necessary to build reliable quantum computers using unreliable components. Quantum computation can be performed with arbitrary accuracy as long as the error rates of physical gates are below a threshold \cite{AB97:FQC:258533.258579}.
The error thresholds  for several schemes have been estimated, and they range from $O(10^{-5})$ to as high as $3\%$ \cite{Ste97L,Ste97a,Ste03:PhysRevA.68.042322,AGP06:QAT:2011665.2011666,Reichardt06,Knill05nature,AC06:PhysRevLett.98.220502,CDT09:CCS:2011814.2011815}.
%Stabilizer codes to protect quantum information [Got97][CRSS97]
%In these analyses of the error thresholds, assumptions such as interaction between any two qubits  are not possible in real physical architectures; only local operations are allowed.
Many of these analyses of error thresholds make simplifying assumptions, such as allowing interactions between
any two qubits, that are not possible in real physical architectures.
Svore, DiVincenzo, and Terhal designed a two-dimensional qubit layout for quantum computation ~\cite{noise_threshold_fault}, using the concatenated Steane code \cite{Ste96a}.
They assumed that two-qubit gates can be applied only to adjacent qubits, and that qubit movement is done by SWAP gates.
Under these assumptions, they showed that the error threshold of the Steane code is $1.85\times~10^{-5}$.
It decreases by roughly a factor of two due to the locality constraints.
Similar work for the concatenated Bacon-Shor code \cite{Bacon06.PhysRevA.73.012340, scheme_reducing_decoherence} was studied by  Spedalieri and Roychowdhury~\cite{latency_local_2d}.
The error threshold reported in~\cite{latency_local_2d} is  also $O(10^{-5})$.

Knill demonstrated a fault-tolerant quantum computation scheme based on concatenated error-detecting codes ($C_4$ and  $C_6$) and postselection with a simulated error threshold of $3\%$
over the depolarizing channel.
Stephens and Evans analyzed a fault-tolerant quantum computation scheme based on the concatenated error-detecting code $C_4$ with locality constraints in one dimension, and
they reported a threshold of $O(10^{-5})$ \cite{SE09_PhysRevA.80.022313}.
In this paper we demonstrate that a two-dimensional layout of the error-detecting code has a
significantly better threshold.
To this end,  we design the optimal qubit movements required to perform quantum computation in a two-dimensional architecture for the concatenated error-detecting code $C_4$  with postselection~\cite{Knill05nature,quantum_accuracy_threshold},
which has the highest known error threshold without locality constraints.
We embed one logical qubit in a $5\times 5$ qubit tile layout.
Our tile has a recursive structure, meaning that each qubit is embedded in a $5\times 5$ tile consisting of lower-level qubits.
As in~\cite{noise_threshold_fault,latency_local_2d}, we assume that two-qubit gates can only be performed locally on adjacent  qubits, and that additional SWAP gates are needed to move the qubits that are far apart.
Each tile contains not only the physical qubits required to maintain the state of a single logical qubit, but also dummy qubits to aid qubit movement by SWAP gates and ancilla qubit preparation for error detection.
In this paper we demonstrate only the tile operations of the error detection block; tile operations for the other gates are available online at

{
\begin{verbatim}
http://mizar.usc.edu/~tbrun/Data/KnillTileOps/
\end{verbatim}
}
%To estimate the error threshold,
%we first use the method of counting malignant pairs in the extended rectangle of the CNOT gate in a local adversarial stochastic noise model \cite{quantum_accuracy_threshold}.
We use both analytical and simulation methods to estimate the error threshold.
The analytical method counts malignant pairs  in the extended rectangle of the CNOT gate in a local adversarial stochastic noise model \cite{quantum_accuracy_threshold}.
In a local adversarial stochastic noise model, arbitrary Pauli errors can be chosen to  attack a given set of gates and we may consider the error threshold obtained from this model to be the lower bound on the threshold for a more realistic error model.
We calculate the thresholds for different ratios of memory error rate to the worst gate error rate.
Assuming that all gates have the same error rates, we obtain a threshold of $3.06\times 10^{-4}$ in a local adversarial stochastic noise model, which is the highest known error threshold for concatenated codes in 2D.

Our second method estimates the threshold by a Monte Carlo simulation of the 2D architecture with depolarizing noise.
We calculate a pseudo-threshold of about $0.1\%$.
%Then we run Monte Carlo simulation of the extended rectangle of the CNOT gate over depolarizing noises to calculate pseudo-thresholds.
As expected, the pseudo-thresholds are generally higher than the thresholds obtained in adversarial noise models.
By setting the memory error rate to be one-tenth of the worst gate error rate,
the error threshold with the adversarial noise model is $4.06\times 10^{-4}$, while the pseudo-threshold with depolarizing noise is about $0.2\%$.

%All quantum technologies have locality requirements --
%For qubits that are far apart, SWAP operation (often constructed from more basic gates such as the CNOT gate) or ballistic movement (in case of ion traps) must be used.
%This movement has an impact on reliability. S

%showed that the error correction threshold is reduced due to movement by a factor of two assuming an optimal placement of the qubits. Since movement reduction is clearly desirable, and %different error correction codes have different structure, we use a \emph{custom qubit layout for each of the three concatenated error correcting codes.}
%For Steane code, we use the optimal qubit layout introduced by Svore et al.~\cite{noise_threshold_fault}, and for the Bacon-Shor code the optimal layout of Since there is no known optimal %layout that reduces movement for the Knill $C_4/C_6$ error correcting code, we designed our own.

% Roadmap
This paper is organized as follows.
In the next section, we describe basic properties of the Knill postselection scheme, and the circuits used to obtain
a universal gate set, using an ancilla factory model.
In the ancilla factory model, ancillary quantum states are distilled so that the phase and the $\pi/8$ gates can be executed.
In Section \ref{sec:2d layout}, we describe the $5\times 5$ two-dimensional qubit tile, and also give the recursive relations of each gate operation in terms of the lower-level gates.
We establish the error threshold by the method of counting the number of malignant pairs in Section \ref{sec:error threshold}.
Simulated pseudo-thresholds are also calculated.
%In Section \ref{sec:overhead}, we calculate the probability of success of the two-dimensional Knill scheme and estimate the overhead of the ancilla factories.
We conclude in Section \ref{sec:conclusion},
including an estimate the error thresholds of a $4\times 4$ tiled qubit layout, obtained  by shrinking the original $5\times 5$ tile.
This tile outperforms the $5\times 5$ tile when memory errors are negligible.

\section{Basics of the Knill $C_4/C_6$ Scheme with Postselection}
\label{Errors_Knill}

% Section overview
%Now we analyze the resource estimates of the quantum computer using the  Knill $C_4/C_6$ code \cite{Knill05nature,quantum_accuracy_threshold}.
%We review some basics of the Knill $C_4/C_6$ scheme with post selection and its fault-tolerant quantum circuits \cite{Knill05nature,quantum_accuracy_threshold}.
In his original scheme, Knill concatenated two error-detecting codes $C_4$ and $C_6$ which alternate.
We follow the simpler version, using only the $C_4$ code as in~\cite{quantum_accuracy_threshold}, which has a high error threshold.
In addition, we concatenate $M$ levels of the quantum error-detecting code $C_{4}$ with a quantum error-correcting code $C_{ec}$ at the top-level.
We use the notation $C_{4}^m$ to denote the Level-$m$ encoding of the $C_{4}$ code
and  the notation $ U_{(m)}$ to denote the gate operation $U$ of $C_4^m$.
We use the notation $\ket{\overline{v}}$ to denote the state $\ket{v}$ at a higher-level of encoding.

The  quantum error-detecting code $C_{4}$ belongs to the class of stabilizer codes \cite{Got97,CRSS97} and can be defined by the stabilizer group with $2$ generators $XXXX$ and $ZZZZ$,
where $X=\begin{pmatrix}0&1\\1&0\end{pmatrix}$ and $Z=\begin{pmatrix}1&0\\0&-1\end{pmatrix}$ are Pauli matrices. %in the computational basis  $\{\ket{0},\ket{1}\}$.
The matrix representation of a single-qubit operator is shown in the computational basis $\{\ket{0},\ket{1}\}$.
This code encodes two logical qubits in four physical qubits and can simultaneously detect any single-qubit bit-flip error $X$
and any single-qubit phase-flip error $Z$.
However, in Knill's scheme, we use only one of the logical qubits and treat the other as a spectator qubit.
The logical operators are
%\begin{align*}
%X^L&=XXII,\\
%Z^L&=ZIZI,\\
%X^S&=IXIX,\\
%Z^S&=IIZZ,
%\end{align*}
$X^L=XXII$,
$Z^L=ZIZI$,
$X^S=IXIX$,
and $Z^S=IIZZ$,
where the superscripts $L$ and $S$ are labels for the logical and the spectator qubits, respectively.
%Thus \[Y^L=iX^L Z^L=YXZI.\]

% Concatenation
The top-level quantum error-correcting code $C_{ec}$ can be the Steane code \cite{Ste96a} or the  Bacon-Shor code \cite{Bacon06.PhysRevA.73.012340, scheme_reducing_decoherence}.
We use the tiled qubit architecture of these two codes studied in~\cite{local_fault_tolerant,latency_local_2d}
on top of the tiled qubit architecture of the $C_4^M$ code developed in the next section.
%The implementation of the logical operations c
%We choose the Steane code in this paper, since the top-level error-correcting code does not affect the error threshold as will be shown in Sec. \ref{sec:error threshold}.
%The analysis of choosing the Bacon-Shor code as the top-level error-correcting code can be similarly developed.
%The Steane code is concatenated in the top-level.
%The logical $\Ket{\bar{0}}$ of the concatenation of $C_{ed}^m$ and $C_{ec}$ is generated by the circuit in Fig. \ref{fig:StatePrepartion4},
%where $\ket{q_1}=\ket{q_2}=\ket{q_4}=\ket{+}_L$ and $\ket{q_3}=\ket{q_5}=\ket{q_6}=\ket{q_7}=\ket{0}_L$.
%The logical $\Ket{\bar{+}}$ at the top level can be similarly generated by switching the control and target qubits of each CNOT gate and choosing
%$\ket{q_1}=\ket{q_2}=\ket{q_4}=\ket{0}_L$ and $\ket{q_3}=\ket{q_5}=\ket{q_6}=\ket{q_7}=\ket{+}_L$.
% Notation
%We will use the same notation defined in \ref{Errors}.
%\begin{figure}[h]
%\centering
%\includegraphics[width=.45\textwidth]{figs/statePrep4}
%\caption{State preparation.}
%\label{fig:StatePrepartion4}
%\end{figure}
We can use the Steane or Shor error correction method, or we can use Knill's syndrome extraction in Fig. \ref{fig:KnillSyndromeExtraction} at the top-level of concatenation.
This choice does not affect the error threshold of the scheme.
In Knill's syndrome extraction, if an error is detected at any error detection step at any level of concatenation, the preparation of the logical EPR pair $\Ket{\overline{\Phi_+}}=\frac{\ket{\overline{00}+\overline{11}}}{2}$ should be restarted.

\begin{figure}[h]
\centering
\[ \Qcircuit @C=0.6em @R=0.3em {
    \lstick{\ket{{Q}}} &\qw   &  \qw                          & \qw      & \qw      & \ctrl{1} & \meterX  & \cw       &\control \cw  &  \\
    \lstick{\ket{{A}}} &\qw& \gate{P_{\ket{\overline{+}}}} & \ctrl{1} & \qw & \targ    & \meterZ  & \control \cw& \\
 \lstick{\ket{{B}}}    &\qw& \gate{P_{\ket{\overline{0}}}} & \targ    &\qw       & \qw      &\qw       &\gate{\overline{X}}\cwx &\gate{\overline{Z}}\cwx[-2] &\qw  & \qw \gategroup{2}{3}{3}{4}{.7em}{_\}} & \rstick{\ket{Q}}\\
                       & &  & \dstick{\text{\small Preparing $\ket{\overline{\Phi_+}}$ } }                      &&&&\\
                       \\
                       \\
} \]
\fcaption{Knill Syndrome extraction .}
\label{fig:KnillSyndromeExtraction}
\end{figure}

%Now we describe the basic fault-toleration circuits of the logical operations of the $C_4$ code to implement
%a universal gate set, using an ancilla factory model to
Now we describe the basic fault-tolerant logical circuits of the $C_4$ code.
We use an ancilla factory model to
prepare the high quality ancillas required to execute the phase gate $S=\begin{pmatrix} 1&0\\0&i\end{pmatrix}$ and the $\pi/8$ gate $T=\begin{pmatrix}1&0\\0&e^{i\pi/4}\end{pmatrix}$.
These gates are then performed by a logical teleportation circuit.

The logical states $\ket{\bar{0}} = \Ket{\bar{0}}_{L} \Ket{\bar{+}}_S$ and $\ket{\bar{+}} = \Ket{\bar{+}}_{L} \Ket{\bar{0}}_S$ can be fault-tolerantly prepared by choosing appropriate spectator qubits as in Fig. \ref{fig:statePrep3},
where $P_{\ket{\bar{0}}}$ and $P_{\ket{\bar{+}}}$   denote the preparation circuits of the logical qubit
$\ket{\bar{0}}$ and $\ket{\bar{+}}$, respectively.
%Note that the state of the spectator qubit determines

%The encoded state $\Ket{\bar{0}}_{L} \Ket{\bar{+}}_S$ and $\Ket{\bar{+}}_L \Ket{\bar{0}}_S$ %correspond to the logical $\Ket{\bar{0}}$ and logical $\Ket{\bar{+}}$, respectively.
%% What is the Steane Code
%appear
%and they can be fault-tolerantly prepared by the circuits in Fig. \ref{fig:statePrep3}.

To perform fault-tolerant error detection (ED) of $C_{4}$, the two circuits in Fig. \ref{fig:FTQEDcircuit} are used depending on the state of the spectator qubit:
we choose ED$_0$ or ED$_+$ when the spectator qubit is $\ket{\bar{+}}_S$ or $\ket{\bar{0}}_S$, respectively.
This is because the state of the spectator qubit alternates between $\ket{\bar{+}}_S$ and $\ket{\bar{0}}_S$ after each error detection block.
As discussed in~\cite{Knill05nature,quantum_accuracy_threshold}, the {ED}$_0$ gate is better suited for detecting $Z$ errors, while the {ED}$_+$  gate is better suited for detecting $X$ errors.

If the parity of the X or Z measurement outcomes in ED$_0$ and ED$_+$ is not zero,
%As long as the parity of the outcome of the $X$ measurements or the parity  of the outcomes of the $Z$ measurements is not zero,
which means that errors are detected, the ancilla qubits are discarded and the circuit restarts.
If there are no errors detected, the measurement outcomes of the the first two code blocks determine the logical Pauli operators to be applied to the second ancilla block to complete
the quantum  teleportation. These operations are represented by the decision block in Fig.  \ref{fig:FTQEDcircuit}.

Each single-qubit gate other than measurements is followed by an ED routine,
and the two-qubit CNOT gate is followed by an ED on each of the two qubits.
As a general rule we shall assume the presence of the input and output error detection routines before and after every logical gate, and this should be understood for every circuit shown.
%For example, $U_{(m)}$ is implemented as the sequence $\left.\text{ED}_{(m)}\right|_{\text{in}} U_{(m)} \left.\text{ED}_{(m)}\right|_{\text{out}}$.
%This sequence is called the $m$-\emph{extended rectangle} ($m$-exRec) of $U$.
%Fig. 12. Illustration of the Steane-EC syndrome extraction method.
%Moreover, when one has multiple gates in action $A_{(m)}B_{(m)}\cdots$,
%each gate is assumed to be of the above form, i.e.,
%$\left.\text{ED}_{(m)}\right|_{\text{in}} A_{(m)} \left.\text{ED}_{(m)}\right|_{\text{out}}\left.\text{ED}_{(m)}\right|_{\text{in}} B_{(m)} \left.\text{ED}_{(m)}\right|_{\text{out}}\cdots.$
%Since ED routines acting back-to-back are only detrimental,
%i.e. same error detection effect but using more gates, they can be contracted yielding
%$\left.\text{ED}_{(m)}\right. A_{(m)} \text{ED}_{(m)} B_{(m)} \left.\text{ED}_{(m)}\right.\cdots.$
%Notice that the EDs in a sequence alternate between ED$_+$ and ED$_0$.
%While this contraction process can be executed for every level of
%concatenation below the one analyzed in order to reduce
%the number of gates, here we opt to contract only at the analyzed level.
Measurements have  quantum ED routines at the input, but classical ED routines at the output,
while ancilla preparations typically have only quantum ED routines at the output.
The combination of a gate and its following ED(s) is called a \emph{rectangle} (1-Rec).
\begin{figure}[h]
\centering
{\footnotesize
\[ \Qcircuit @C=0.4em @R=1.3em {
&&&&&& &\lstick{\ket{+}}&\qw &\ctrl{2} &\qw &\qw&\qw &&&&&&&&&&&&&&&&&&&&&&&&\\
\lstick{P_{\ket{\overline{0}}}}&&&&&& &\lstick{\ket{+}}&\qw & \qw  &\ctrl{2} &\qw&\qw&\rstick{\ket{0}_L\ket{+}_S}&&\\
&&&&&& &\lstick{\ket{0}}&\qw & \targ &\qw &\qw&\qw&&&&&\\
&&&&&& &\lstick{\ket{0}}&\qw& \qw & \targ&\qw \gategroup{1}{2}{4}{2}{0.7em}{\{}&\qw \gategroup{1}{13}{4}{13}{0.7em}{\}}
 }
 \Qcircuit @C=0.3em @R=1.3em {
&&&&&&&& &\lstick{\ket{+}}&\qw &\ctrl{1} &\qw & \qw &&&\\
\lstick{P_{\ket{\overline{+}}}}&&&&&&&& &\lstick{\ket{0}}&\qw & \targ &\qw&\qw&&&\rstick{\ket{+}_L\ket{0}_S}\\
&&&&&&&& &\lstick{\ket{+}}&\qw & \ctrl{1} &\qw &\qw \\
&&&&&&&& &\lstick{\ket{0}}&\qw&  \targ&\qw \gategroup{1}{2}{4}{2}{0.7em}{\{} &\qw \gategroup{1}{14}{4}{14}{0.7em}{\}}\\
 }
  \]
}
\fcaption{State preparation .}
\label{fig:statePrep3}
\end{figure}
\begin{figure}[h]
\centering
\[ \Qcircuit @C=0.2em @R=0.3em {
    \lstick{\ket{{\Psi}}_L} &{/}\qw  &\qw &  \qw                          & \qw      & \ctrl{1} & \meterX  &  \control  \cw &  \\
\lstick{\ket{{0}}_L\ket{+}_S}&{/}\qw& \qw&\gate{P_{\ket{\overline{+}}}} & \ctrl{1} & \targ    & \meterZ    & \control \cw \cwx\\
\lstick{\ket{{+}}_L\ket{0}_S}&{/}\qw& \qw&\gate{P_{\ket{\overline{0}}}} & \targ    & \qw      &\qw       &\gate{\text{\scriptsize Decision}}\cwx &\qw & {/}\qw & \qw &\qw &\rstick{ \ket{\Psi}_L\ket{0}_S}
} \]
\[ \Qcircuit @C=0.2em @R=0.3em {
    \lstick{\ket{{\Psi}}_L} &{/}\qw  &\qw &  \qw                          & \qw      & \ctrl{1} & \meterX  &  \control  \cw &  \\
\lstick{\ket{{+}}_L\ket{0}_S}&{/}\qw& \qw&\gate{P_{\ket{\overline{+}}}} & \ctrl{1} & \targ    & \meterZ    & \control \cw \cwx\\
\lstick{\ket{{0}}_L\ket{+}_S}&{/}\qw& \qw&\gate{P_{\ket{\overline{0}}}} & \targ    & \qw      &\qw       &\gate{\text{\scriptsize Decision}}\cwx &\qw & {/}\qw & \qw &\qw &\rstick{\ket{\Psi}_L\ket{+}_S}
} \]
\fcaption{Circuits for fault-tolerant quantum error detection.}{\footnotesize Top: {ED}$_0$. Bottom: {ED}$_+$.}
\label{fig:FTQEDcircuit}
\end{figure}

The logical controlled-NOT (CNOT) gates between different code blocks of $C_4$ can be done  transversally
by applying  bitwise CNOT gates.
The swap of qubits 2 and 3 implements the SWAP gate of the logical qubit and the spectator qubit, and we call this an \emph{inner} SWAP gate.
%, which is opposed to the \emph{outer} SWAP between two code blocks.
The logical Hadamard gate $H=\frac{1}{\sqrt{2}}\begin{pmatrix}1&1\\1&-1\end{pmatrix}$ is implemented by transversally applying the Hadamard gates, followed by an inner SWAP gate.
The inner SWAP gate does not need to be applied; instead, we switch the labels of the qubits and keep track of them.
We assume this can be done efficiently.

To enable universal quantum computation, %we need to implement the phase gate $S$ and the $\pi/8$ gate.
%Unfortunately, a fault-tolerant version of these two gates cannot be constructed transversally.
it remains to prepare the level-$M$ ancilla state $\Ket{\overline{+i}}=\frac{1}{\sqrt{2}}\left(\Ket{\overline{0}}+i\Ket{\overline{1}} \right)$, which is the $+1$ eigenstate of $Y=iXZ$ at level $M$,
%the magic state $\Ket{H}=\cos\left( \frac{\pi}{8}\right)\Ket{0}+\sin\left( \frac{\pi}{8}\right)\Ket{1}$, the $+1$ eigenstate of $H$.
and the level-$M$ magic state $T\Ket{\bar{+}}$.
%We use the ancilla factory of the Bacon-Shor code as described in the previous section.
%The only difference is the decoding operation in the injection circuit,
%which is shown in Fig. \ref{fig:decoding_Ced} for the $C_4$ code.
The phase gate $S$ and the $\pi/8$ gate $T$  can be implemented with the help of the ancilla state $\Ket{\overline{+i}}$ and $T\Ket{\bar{+}}$ as shown in Fig. \ref{fig:Knill_phase_gate} and Fig. \ref{fig:tgate}, respectively.

The logical state $\ket{\overline{+i}}$  can be non-fault-tolerantly prepared  by the circuit in Fig. \ref{fig:state_i_prep}.
To prepare the physical state $\ket{+i}= SH \Ket{0}$ at level 0, we sequentially apply the faulty gates $H$ and $S$ on a physical qubit $\Ket{0}$.
After several iterations of distillation, we obtain a $\Ket{{+i}}$ with high fidelity.
The decoding gate $\mathcal{D}$ is shown in Fig. \ref{fig:decoding_Ced}.
The output state $\Ket{\overline{+i}}$ can be distilled to one with higher fidelity by the circuit in Fig. \ref{fig:distill},
where the twirl operation is shown in Fig. \ref{fig:twirl}.
The state $\Ket{\overline{+i}}$ at level $M$ can be prepared by recursively applying the circuit in Fig. \ref{fig:state_i_prep}
or by using a level-$M$ to level-$0$ decoder $\mathcal{D}$ in the teleportation at level $M$.
A level-$M$ to level-$0$ decoder can be implemented by recursively applying the decoding gate $\mathcal{D}$ at each level.

%The magic state $\Ket{H}$ is prepared by distilling given faulty $\Ket{H}$ to be with higher fidelity.

\begin{figure}[h!]
\centering
\[ \Qcircuit @C=0.5em @R=1.5em {
    \lstick{\ket{\overline{\Psi}}} &\qw & \ctrl{1} & \ctrl{1} & \qw &  \qw&  \rstick{S\ket{\overline{\Psi}}} \\
   \lstick{\ket{\overline{+i}}} & \qw& \targ & \gate{Z}  & \qw &\qw &\rstick{\ket{\overline{+i}}}
} \]
\fcaption{The circuit for implementing the logical $S$ gate.}
\label{fig:Knill_phase_gate}
\end{figure}

\begin{figure}[h!]
\centering
\[ \Qcircuit @C=0.5em @R=1.5em {
    \lstick{\ket{\overline{\Psi}}} & {/}\qw &  \qw & \targ & \meterZ & \control \cw&  \cw\\
   & {/}\qw & \gate{T\ket{\overline{+}}} & \ctrl{-1} & \qw  & \gate{\overline{S}} \cwx & \gate{\text{EC}}& \qw & \rstick{T\ket{\overline{\psi}}}
} \]
\fcaption{The circuit for implementing the logical T gate.
  }
\label{fig:tgate}
\end{figure}

\begin{figure}[h!]
\centering
\[ \Qcircuit @C=0.4em @R=0.3em {
    \lstick{\ket{{+i}}} &\qw  &\qw &  \qw                          & \qw      & \qw      & \ctrl{1} & \meterX  & \cw         &\control  \cw &  \\
                        &{/}\qw& \qw&\gate{P_{\ket{\overline{+}}}} & \ctrl{1} & \gate{D} & \targ    & \meterZ  & \control \cw& \cwx\\
                        &{/}\qw& \qw&\gate{P_{\ket{\overline{0}}}} & \targ    &\qw       & \qw      &\qw       &\gate{\overline{X}}\cwx &\gate{\overline{Z}}\cwx{2}&\qw & {/}\qw & \qw &\qw &\rstick{\ket{\overline{+i}}}
} \]
\fcaption{The circuit for preparing the logical state $\Ket{\overline{+i}}$.}
\label{fig:state_i_prep}
\end{figure}
\begin{figure}[h!]
\centering
{\footnotesize
\[ \Qcircuit @C=0.4em @R=1.0em {
&&&&&& &\lstick{\ket{q_1}}&\qw &\targ &\qw &\ctrl{1}&\qw &\rstick{\ket{\Psi}}&&&&&&&&&&&&&&&&&&&\\
\lstick{\ket{\Psi}_L\ket{0}_S}&&&&&& &\lstick{\ket{q_2}}&\qw & \qw  &\targ &\targ&\qw&&&&&\\
&&&&&& &\lstick{\ket{q_3}}&\qw & \ctrl{-2} &\qw &\qw&\qw&&&&&\\
&&&&&& &\lstick{\ket{q_4}}&\qw& \qw & \ctrl{-2}&\qw&\qw \gategroup{1}{2}{4}{2}{0.7em}{\{}&&&&&
 }
 \Qcircuit @C=0.3em @R=1.0em {
&&&&&&&& &\lstick{\ket{q_1}}&\qw &\ctrl{1} &\qw &\targ& \qw &\qw&\rstick{\ket{\Psi}}\\
\lstick{\ket{\Psi}_L\ket{+}_S}&&&&&&&& &\lstick{\ket{q_2}}&\qw & \targ &\qw&\qw&\qw&\qw\\
&&&&&&&& &\lstick{\ket{q_3}}&\qw & \ctrl{1} &\qw &\ctrl{-2}&\qw&\qw\\
&&&&&&&& &\lstick{\ket{q_4}}&\qw&  \targ&\qw&\qw &\qw&\qw \gategroup{1}{2}{4}{2}{0.7em}{\{}
 }
  \]
}

\fcaption{The decoding circuit for $C_{4}$.}
\label{fig:decoding_Ced}
\end{figure}
\begin{figure}[h!]
\centering
\[ \Qcircuit @C=0.5em @R=1.5em {
    \lstick{\ket{\overline{+i}}} & \gate{\text{twirl}} & \ctrl{1} & \targ & \qw &  \qw&  \rstick{\ket{\overline{+i}}} \\
   \lstick{\ket{\overline{+i}}} & \gate{\text{twirl}} & \gate{Z} & \ctrl{-1} & \meterX  & \cw & \cw & \control \cw
} \]
\fcaption{The distillation circuit for the state $\Ket{+i}$.}
\label{fig:distill}
\end{figure}
\begin{figure}[h!]
\centering
\[ \Qcircuit @C=0.5em @R=1.5em {
    \lstick{\ket{+}}          &\qw & \meterZ &  \cw & \control \cw  & \cw \\
   \lstick{\ket{\overline{+i}}} &\qw & \qw   & \qw  & \gate{Y}\cwx & \qw & \rstick{\ket{\overline{+i}}} \qw
} \]
\fcaption{The twirl operation for the state $\Ket{+i}$.}
\label{fig:twirl}
\end{figure}

%\begin{figure}[h]
%\centering
%\includegraphics[width=.45\textwidth]{figs/KnillErrorCorrection}
%\caption{Knill Error Correction.}
%\label{fig:KnillErrorCorrection}
%\end{figure}
%

% The T gate
%In order to have a universal gate set, we also need the $\pi/8$ gate called the $T$ gate.
The realization of a fault-tolerant $T$ gate is shown in Fig.~\ref{fig:tgate}.
This gate sequence was originally constructed in~\cite{methodology_quantum_logic} using one-bit teleportation.
The gate sequence teleports the state $\Ket{\psi}$ from the data block to the ancilla and applies the $T$ gate to the state.
The ancilla state $T \ket{\bar +}$ is prepared using the state injection method described before, as in Fig. \ref{fig:state_i_prep}, followed by several rounds of distillation.
%   This state is injected via the same circuit as Fig. \ref{fig:state_i_prep}.
The distillation and twirl procedures of $T\ket{+}$ are complicated and they are %different from those of the $\ket{+i}$ state as
described in~\cite{BK05_PhysRevA.71.022316}.

\section{The Two-Dimensional Qubit Layout of the $C_4$ Code}
\label{sec:2d layout}
%\subsection{The Qubit Layout of the $C_4$ Code}
We now describe the two-dimensional  qubit layout for the $C_4$ code and estimate the number of each physical gate operation required for each logical operation. % and  their execution time steps with parallelism taken into account.
For examples of estimation of the resources of the Knill scheme based on the concatenated  $C_4$ code, we refer interested readers to our technical report \cite{QCS_tech12}.
We assume that  two-qubit interactions are available only for the nearest neighbors.
That is, we apply horizontal or vertical CNOT gates (hCNOT/vCNOT) only to two neighboring qubits on the same horizontal or vertical line.
Similarly, we assume movements of the qubits are accomplished by SWAP gates in two directions: horizontal and vertical SWAP gates (hSWAP/vSWAP).
%Let $ops( \text{Gate}_{(M+1)})$  denote the gate operation of the concatenated code at level $M+1$ ($M$ levels of $C_{ed}$ and one level of $C_{ec}$),
%The recursive relations of each $ops( \text{Gate}_{(M+1)})$ in terms of the  $ops( \text{Gate}^{C_4}_{(M)})$'s are listed in Table \ref{table:recursive gate counts}.
%Similarly, let $time(\text{Gate}_{(m)}))$ and $time(\text{Gate}^{C_4}_{(m)}))$ denote the execution time required and the recursive relations of
%the required time are listed in Table \ref{table:recursive gate time}.
%A gate $U$ at level $m$, $U(m)$, is followed by an error detection routine at level $m$, ED$(m)$.
%The combination of $U(m)$ and its following ED$(m)$ is called an $m$-rectangle ($m$-Rec) of $U$.
%Let $ U_{(m)}$  denote the gate operation $U$ of the $C_4$ code at level $m$.

Following the tile structures presented in~\cite{local_fault_tolerant,latency_local_2d},
we design a two-dimensional  $5\times 5$  lattice architecture of physical qubits to represent a logical qubit of the $C_4$ code.
A tile is initialized as one of the following two structures:
{\scriptsize
\begin{align*}
\text{Structure } \textrm{I}:
\left[
\begin{array}{cccccccccc}
O& & O  & & O & & O & & O\\
 & &    & &   & &   & &   \\
O& & d_1& & O & & O & & d_3\\
 & &    & &   & &   & &   \\
O& & O  & & O & & O & & O\\
 & &    & &   & &   & &   \\
O& & O  & & O & & O & & O\\
 & &    & &   & &   & &   \\
O& & d_2& & O & & O & & d_4
\end{array}\right],
\end{align*}
}
{\scriptsize
\begin{align*}
\text{Structure } \textrm{II}:
\left[\begin{array}{cccccccccc}
O& & O  & & O & & O & & O\\
 & &    & &   & &   & &   \\
O& & O& & O & & O & & O\\
 & &    & &   & &   & &   \\
O& & O  & & d_1 & & d_3 & & O\\
 & &    & &   & &   & &   \\
O& & O  & & d_2 & & d_4 & & O\\
 & &    & &   & &   & &   \\
O& & O & & O & & O & & O
\end{array}\right].
\end{align*}
}
The four data qubits of the $C_4$ code are denoted by $d_1$, $d_2$, $d_3$, and $d_4$.
The $O$'s are dummy qubits used for ancilla preparation  or for swapping with data or ancilla qubits in communication, and
their states are irrelevant to computation.
Each qubit in the tile is encoded in a lower-level tile structure.
%The dummy qubits

The following operations are performed on the C4 code:
\begin{enumerate}
  \item Error detection (ED).
  \item Horizontal and vertical CNOT gates (hCNOT/vCNOT).
  \item Horizontal and vertical SWAP gates (hSWAP/vSWAP).
  \item Measurement in the X basis or the $Z$ basis  ($M_X$ and $M_Z$).
  \item The Pauli operators $X$, $Y$, $Z$, and the Hadamard gate ($H$).
  \item Preparation of the ancilla qubits $\ket{+}$ or $\ket{0}$ ($P_{{\ket{+}}}$ and $P_{\Ket{0}}$).
%  \item decoding circuit
  \item The phase gate $S$ and  the $\pi/8$ gate $T$.
\end{enumerate}
%the error detection circuit $ED_+$ and $ED_0$ in Fig. \ref{fig:FTQEDcircuit}.
For simplicity, all lower-level gates are assumed to take one \emph{time step}, which is the longest execution time among all gates.
In reality, we may think that a qubit idles for the rest of the time step after it completes a fast gate,  and the error rate of this operation
is the physical gate error rate plus the memory error rate for the idle time.
%We try to optimize the gate operations in the numbers of SWAPs, idle qubits, and time steps.
To achieve favorable error-correction
properties and low overhead, our design attempts to minimize the number of SWAPs, idle qubits, and the total number of time steps.

Since error detection is performed constantly, extra space is need for preparation of the logical EPR pairs used in Knill's syndrome extraction.
In structure \textrm{I}, the data qubits lie on the ``corners" and the logical EPR pairs for error detection are prepared inside the data qubits.
By contrast, the data qubits are located at the ``center" of structure \textrm{II} and the ancilla qubits surround the data qubits, as will be shown in the following.
Error detection is designed in each of the two tiles precisely so that the data qubits are transferred between  the ``center" and the ``corners."
Therefore, the tile alternates between structures \textrm{I} and \textrm{II} after each error detection.
This alternation avoids extra SWAP gates.
%We use the tile of the Steane code at the top-level of concatenation.

  For a SWAP operation to be fault-tolerant, we only swap a data or ancilla qubit with a dummy qubit.
  Because of this, only one tile requires an ED circuit.
%For realization of the horizontal or vertical CNOT gate,
%the first row and the first column of the tile are preserved for transportation of the data qubits
%such that constantly EDs are not affected.
The topmost row and leftmost column of each tile is reserved for transportation of the lower level qubits.
This allows realization of the horizontal or vertical CNOT gate without affecting the EDs.

%To save space, detailed tile operations are available online.
%The dummy qubits enclosed by the four data qubits are elegantly designed for error detection depending on the state of the spectator qubit.
%The dummy qubits on the first row and the first column of the tile are preserved for qubit movements.
%Herein we only demonstrate the ED$_+$ block in Fig. \ref{fig:FTQEDcircuit} for structure \textrm{I}.
Fig. \ref{fig:FTQEDcircuit} demonstrates the ED$_{+}$ block for structure \textrm{I}.
Due to space constraints, we have made the other tile operations available online.
In the following, $``a \rightarrow b"$  means applying a CNOT gate with $a$ being the control qubit  and  $b$ being the target qubit.\\

\noindent Time step 1:
  {\footnotesize
\begin{align*}
\begin{array}{cccccccccc}
O& & O  & & O & & O & & O\\
 & &    & &   & &   & &   \\
O& & d_1& & O & & O & & d_3\\
 & &    & &   & &   & &   \\
O& & P_{\Ket{+}}(a_1)  & & P_{\Ket{+}}(a_5)  & & P_{\Ket{0}}(a_7)  & & P_{\Ket{+}}(a_3) \\
 & &    & &   & &   & &   \\
O& & P_{\Ket{0}}(a_2)  & & P_{\Ket{+}}(a_6)  & & P_{\Ket{0}}(a_8)  & & P_{\Ket{0}}(a_4) \\
 & &    & &   & &   & &   \\
O& & d_2& & O & & O & & d_4
\end{array}
\end{align*}
}
Time step 2:
{\footnotesize
\begin{align*}
\begin{array}{cccccccccc}
O& & O  & & O & & O & & O\\
 & &    & &   & &   & &   \\
O& & d_1& & O & & O & & d_3\\
 & &    & &   & &   & &   \\
O& & a_1  & & a_5  &\rightarrow & a_7  & & a_3 \\
 & & \downarrow   & &   & &   & & \downarrow   \\
O& & a_2  & & a_6  &\rightarrow & a_8  & & a_4 \\
 & &    & &   & &   & &   \\
O& & d_2& & O & & O & & d_4.
\end{array}
\end{align*}
}
Time step 3:
{\footnotesize
\begin{align*}
\begin{array}{cccccccccc}
O& & O  & & O & & O & & O\\
 & &    & &   & &   & &   \\
O& & d_1& & O & & O & & d_3\\
 & &    & &   & &   & &   \\
O& & a_1  &\rightarrow & a_5  & & a_7  & \leftarrow& a_3 \\
 & &    & &   & &   & &   \\
O& & a_2  & \rightarrow& a_6  & & a_8  & \leftarrow& a_4 \\
 & &    & &   & &   & &   \\
O& & d_2& & O & & O & & d_4
\end{array}
\end{align*}
}
Time step 4:
{\footnotesize
\begin{align*}
\begin{array}{cccccccccc}
O& & O  & & O & & O & & O\\
 & &    & &   & &   & &   \\
O& & d_1& & O & & O & & d_3\\
 & & \downarrow   & &   & &   & & \downarrow  \\
O& & a_1  & & a_5  & & a_7  & & a_3 \\
 & &    & &   & &   & &   \\
O& & a_2  & & a_6  & & a_8  & & a_4 \\
 & & \uparrow   & &   & &   & &  \uparrow \\
O& & d_2& & O & & O & & d_4
\end{array}
\end{align*}
}
Time step 5:
{\footnotesize
\begin{align*}
\begin{array}{cccccccccc}
O& & O  & & O & & O & & O\\
 & &    & &   & &   & &   \\
O& & M_X(d_1)& & O & & O & & M_X(d_3)\\
 & &    & &   & &   & &   \\
O& & M_Z(a_1)  & & a_5  & & a_7  & & M_Z(a_3) \\
 & &    & &   & &   & &   \\
O& & M_Z(a_2)  & & a_6  & & a_8  & & M_Z(a_4) \\
 & &    & &   & &   & &   \\
O& & M_X(d_2)& & O & & O & & M_X(d_4)
\end{array}
\end{align*}
}
At the end of time step 5:
{\footnotesize
\begin{align*}
\begin{array}{cccccccccc}
O& & O  & & O & & O & & O\\
 & &    & &   & &   & &   \\
O& & O& & O & & O & & O\\
 & &    & & & &   & &   \\
O& & O & & d_1  & & d_3  & & O \\
 & &    & &   & &   & &   \\
O& & O  & & d_2  & & d_4  & & O \\
 & &    & &   & &  & &  \\
O& & O  & & O & & O & & O
\end{array}
\end{align*}
}
The ancilla qubits $a_1,$ $\cdots,$ $a_8$ are prepared at time step 1,   and logical EPR pairs are made at time steps 2 and 3.
Quantum teleportations are completed in the subsequent time steps.
We choose the index such that a quantum teleportation occurs on the qubits $d_i, a_i, a_{i+4}$ for $i=1,2,3,4$.
Observe that the data qubits  $d_1, d_2, d_3, d_4$ are transferred to the center after teleportation
and no SWAPs are needed here.
However, the error detection ED$_+$ for structure \textrm{II} needs two SWAPs and it takes one more step. Its first time step is initialized as follows:
{\footnotesize
\begin{align*}
\begin{array}{cccccccccc}
O& & O  & & O & & O & & O\\
 & &    & &   & &   & &   \\
O& & O& & P_{\Ket{+}}(a_1)  & & P_{\Ket{0}}(a_3)  & & O\\
 & &    & & & &   & &   \\
O& & P_{\Ket{+}}(a_5)  & & d_1  & & d_3  & & P_{\Ket{+}}(a_7)  \\
 & &    & &   & &   & &   \\
O& & P_{\Ket{0}}(a_6)   & & d_2  & & d_4  & & P_{\Ket{0}}(a_8)  \\
 & &    & &   & &  & &  \\
O& & O  & & P_{\Ket{+}}(a_2)  & & P_{\Ket{0}}(a_4)  & & O
\end{array}.
\end{align*}
}
In addition, applying the logical Pauli operators $X$ or $Z$ to complete the teleportation may take one or two more steps, but this is not shown.
In many cases it suffices to track these Pauli operators without correcting them.
The ED$_0$ for structure \textrm{I} at time step 1 is as follows and the rest of the steps are similar to those of the above ED$_+$:
{\footnotesize
\begin{align*}
\begin{array}{cccccccccc}
O& & O  & & O & & O & & O\\
 & &    & &   & &   & &   \\
O& & d_1& & P_{\Ket{+}}(a_1) & & P_{\Ket{0}}(a_3) & & d_3\\
 & &    & &   & &   & &   \\
O& & O  & & P_{\Ket{+}}(a_5)  & & P_{\Ket{+}}(a_7)  & & O \\
 & &    & &   & &   & &   \\
O& & O  & & P_{\Ket{0}}(a_6)  & & P_{\Ket{0}}(a_8)  & & O \\
 & &    & &   & &   & &   \\
O& & d_2& &P_{\Ket{+}}(a_2) & & P_{\Ket{0}}(a_4) & & d_4
\end{array}.
\end{align*}
}
%We found the operations and required time of ED$_0$ are the same as those of ED$_+$
%and we omit the subscripts $0$ or $+$. %, since the circuit for $ED_0$ is similar to that for $ED_+$
Note that the operations and required time of ED$_0$ are the same as those of ED$_+$.

%Let $ops( \text{Gate}_{(M+1)})$  denote the gate operation of the concatenated code at level $M+1$ ($M$ levels of $C_{ed}$ and one level of $C_{ec}$),
%and $ops( \text{Gate}^{C_4}_{(m)})$  denote the gate operation of the $C_4$ code at level $m$.

Remark: after a logical Hadamard gate, the labels of data qubits $2$ and $3$ are switched.
This can be fixed by applying appropriate SWAPs and it takes two more time steps in
structures~\textrm{I} or \textrm{II}.
However, we don't adjust it until a CNOT gate acts on two tiles with different labels.

Based on the tiled operations, we build the recursive relations resulting from
the concatenated code structure in order to quantify the total
number of gates and total time required for each logical gate.
The recursive relations of a 1-Rec for each logical gate in terms of lower-level gates are listed in Table \ref{table:recursive gate counts}.
For example, the vertical SWAP operation vSWAP requires 20 vertical swap operations at the next lower concatenation level, followed by the error detection operation ED.
{
\begin{table*}
\scriptsize
\begin{center}

\begin{tabular}{|c|c|c|c|c|c|c|c|}
%$ops(ED_{m-1}^{C_4})$& $ops(CNOT_{(m-1)}^{C_4})$& $ops(ED_{(m-1)}^{C_4})$& $ops(M_{Z(m-1)}^{C_4})$& $ops(M_{X(m-1)}^{C_4})$&$ ops(P_{\ket{+}(m-1)}^{C_4})$&$0$ &\\
  \hline
   &  ED$_{ (1)}$ & vCNOT$_{(1)}$ & hCNOT$_{(1)}$ & vSWAP$_{(1)}$ & hSWAP$_{(1)}$ &  $P_{\ket{+}(1)}$ & $P_{\ket{0}(1)}$ \\
\hline
ED$_{(1)}$       & & 2& 2& 1& 1& 1 &1\\
vCNOT$_{(0)}$    &6& 4&  & &  & 2 &\\
hCNOT$_{(0)}$    &6&  & 4& &  &  &2\\
vSWAP$_{(0)}$    &*&40& 8&20& & 4 &\\
hSWAP$_{(0)}$    &*& 8&40& &20&  &4\\
$P_{\ket{+}(0)}$ &4&  &  &  & & 2 &2\\
$P_{\ket{0}(0)}$ &4&  &  &  &  &2& 2\\
$M_{Z(0)}$   &4&&&&&  &\\
$M_{X(0)}$   &4&&&&&  &\\
$Z_{(0)}$&&&&&&  &\\
$X_{(0)}$&&&&&&  &\\
$H_{(0)}$&&&&&&  &\\
%$S_{(0)}$&1&0/8&20/4&&&&&&&8\\
%$T_{(0)}$&2&0/12&40/12&&&&4&&&8\\
  \hline
\end{tabular}\\
\begin{tabular}{|c|c|c|c|c|c|c|c|}
%$ops(ED_{m-1}^{C_4})$& $ops(CNOT_{(m-1)}^{C_4})$& $ops(ED_{(m-1)}^{C_4})$& $ops(M_{Z(m-1)}^{C_4})$& $ops(M_{X(m-1)}^{C_4})$&$ ops(P_{\ket{+}(m-1)}^{C_4})$&$0$ &\\
  \hline
   &   $M_{Z(1)}$ & $M_{X(1)}$   &  $Z_{(1)}$ &$X_{(1)}$& $H_{(1)}$&$S_{(1)}$&$T_{(1)}$ \\
\hline
ED$_{(1)}$       &&&1&1&1&1&2\\
vCNOT$_{(0)}$    &&&&&&  &\\
hCNOT$_{(0)}$    &&&&&& 8 &12\\
vSWAP$_{(0)}$    &&&&&& 20 &40\\
hSWAP$_{(0)}$    &&&&&& 4 &12\\
$P_{\ket{+}(0)}$ &&&&&&  &\\
$P_{\ket{0}(0)}$ &&&&&&  &\\
$M_{Z(0)}$   &4&&&&&  &\\
$M_{X(0)}$   &&4&&&&  &4\\
$Z_{(0)}$&&&2&&&  &\\
$X_{(0)}$&&&&2&&  &\\
$H_{(0)}$&&&&&4&8  &8\\
%$S_{(0)}$&1&0/8&20/4&&&&&&&8\\
%$T_{(0)}$&2&0/12&40/12&&&&4&&&8\\
  \hline
\end{tabular}\\
($^{*}$The number of SWAPs in the ED is zero in structure \textrm{I} but $4$ in structure \textrm{II}.)
\end{center}
\tcaption{The numbers of the quantum operations contained in each higher-level quantum operation  of the $C_4$ code and its following error detection (1-Rec).
Each entry represents the number of the elementary gate $(U_{(0)})$ corresponding to that row contained in the higher-level gate ($U_{(1)}$) corresponding to that column.
 }\label{table:recursive gate counts}
\end{table*}
}
To allow universal quantum computation, we implement the $S$ and $T$ gates by the ancilla factory method,
which uses the decoding circuits in Fig. \ref{fig:decoding_Ced}.
The overhead of the ancilla factories and their decoding circuits are not included in Table  \ref{table:recursive gate counts}. % and \ref{table:recursive gate time}

Remark: It is possible to combine a gate operation with the following error detection and save several time steps.

\section{Error Analysis of the 2-Dimensional Knill Postselection Scheme}
\label{sec:error threshold}

\subsection{Error Threshold}
Here we estimate the error threshold of Knill's postselection scheme  in the 2-dimensional tile for the \emph{local stochastic, adversarial noise model}.
Following the procedure presented in \cite{quantum_accuracy_threshold,noise_threshold_fault,latency_local_2d}, we count the number of \emph{malignant pairs}  of {locations} in the 1-exRec of the CNOT gate.
%A rectangle of the CNOT gate .

The 1-exRec of the CNOT gate includes the bitwise CNOTs on two logical qubits, together with two following and two preceding error detection blocks.
%is a rectangle of the CNOT gate with two preceding error detection  blocks.
As shown in Fig. \ref{fig:1-exRec CNOT}, we assume the preceding EDs are ED$_+$s and the following EDs are ED$_0$s.
Note that the logical Pauli operators to complete teleportation  are assumed to be error-free, since they can be tracked in the Pauli frame and hence be deferred until a non-Clifford gate occurs.
We also assume that classical computations are perfect, and that  any quantum operations depending on the classical results can be applied without delay.

\begin{figure}[h]
\centering
\[ \Qcircuit @C=2em @R=1.7em {
   & \gate{\text{ED}_+} & \ctrl{1} & \gate{\text{ED}_0} & \qw \\
   & \gate{\text{ED}_+} & \targ & \gate{\text{ED}_0} & \qw
} \]
\fcaption{The 1-exRec of the CNOT gate.}
\label{fig:1-exRec CNOT}
\end{figure}

There are seven types of locations in the 1-exRec of the CNOT gate:
(1)$P_{\ket{+}}$; (2)$P_{\ket{0}}$;
(3)$M_X$; (4) $M_Z$; (5) hSWAP/vSWAP; (6) hCNOT/vCNOT; (7) idle qubits.
A set of locations is called malignant if errors happening in these locations could make the calculation of the rectangle  incorrect.
Since an error at any single location can be detected, errors at two locations dominate the source of logical errors.
To determine whether a pair of locations is malignant,
we check whether there is any logical error in the output of two perfect ED$_+$s following the 1-exRec of CNOT, as shown in Fig. \ref{fig:1-exRec CNOT_ED+}.
The simulation procedure in the stabilizer formalism proposed in~\cite{Aaronson04.PhysRevA.70.052328} can be used to track the logical operators through the circuit in Fig. \ref{fig:1-exRec CNOT_ED+}.

\begin{figure}[h]
\centering
\[ \Qcircuit @C=2em @R=1.7em {
   & \gate{\text{ED}_+} & \ctrl{1} & \gate{\text{ED}_0} & \qw & \gate{\text{ideal ED}_+}& \qw\\
   & \gate{\text{ED}_+} & \targ & \gate{\text{ED}_0} & \qw & \gate{\text{ideal ED}_+}& \qw \gategroup{1}{6}{2}{6}{0.7em}{--}
} \]
\fcaption{The 1-exRec of the CNOT gate followed by two perfect ED$_+$.}
\label{fig:1-exRec CNOT_ED+}
\end{figure}

Remark: in general the error rate of  a SWAP gate is higher than a CNOT gate, since it is implemented by a series of gate operations, such as three CNOT gates.
However, in the two-dimensional tile we only swap a data or ancilla qubit with a dummy qubit, and the cost of such a SWAP gate is less than the cost of a CNOT gate. %as can be seen in Appendix \ref{apendix:tile_implementation_C4}.
As for the $S$ and $T$ gates, Aliferis, Gottesman, and Preskill showed that the distillation method for ancilla preparations has a higher threshold than the code itself~\cite{quantum_accuracy_threshold}.

To maximize the error threshold, we optimized the tile operations of the extended rectangle of the CNOT gate. The animations showing this are available online.
There are $196$ locations in the extended rectangle of the CNOT gate: $32$ idle qubits and $154$ gates, of which $38$ gates are SWAPs.
We assume that the error detection blocks begin before the time step that the data qubits come in, and thus there are no idle qubits at time steps 1, 2, and 3 in the preceding ED.
We find that the numbers of malignant pairs of locations of each kind are given by
\[
\footnotesize
\alpha= \begin{pmatrix}
4&	8&	8&	0&	0&	32&	16\\
&	0&	0&	14&	96&	80&	32\\
&	&	16&	0&	96&	104&	32\\
&	&	&	16&	96&	112&	32\\
&	&	&	&	442&	672&	268\\
&	&	&	&	&	322&	288\\
&	&	&	&	&	&	106\\
\end{pmatrix},
\]
where $\alpha_{i,j}$ represents the number of malignant pairs at locations of types $i$ and $j$.

Let $\epsilon^{(m)}_j$ be the error rates of type $j$  at level $m$.
For error correction to be  effective, we require
\begin{align} \label{eq:error threshold bound}
\epsilon_{6}^{(m+1)}= \sum_{i\leq j} \alpha_{i,j} \epsilon^{(m)}_{i}\epsilon^{(m)}_{j}+ O((\epsilon^{(m)}_{\text{max}})^3)\leq \epsilon_{6}^{(m)},
\end{align}
where $\alpha_{i,j}$ is the number of malignant pairs of types $i$ and $j$ and $\epsilon^{m}_{\text{max}}$ is the maximum of the seven types of error rate.

We assume all errors of weight 3 or larger are malignant and the effect of errors of weight higher than three can be ignored.
(This might still be an overestimate of higher-order terms.)
Let  $\gamma$ be the ratio of the memory error rate of the idle qubits to the gate error rate.
Let
\[
B= {\sum_{\underset{i>j}{i,j,=1}}^6} \alpha_{i,j}+  \sum_{i=1}^6 \gamma \alpha_{i,7}+ \gamma^2 \alpha_{7,7}
\]
be the effective number of malignant pairs and
\[
A= {164\choose 3}+{164\choose 2}{32\choose 1}\gamma +{164\choose 1}{32\choose 2}\gamma^2+ {32\choose 3}\gamma^3
\]
be the effective number of errors  of weight 3, where the ${a \choose b}$'s are binomial coefficients.
Then Eq. (\ref{eq:error threshold bound}) reduces to
\[
A (\epsilon^{(m)}_{6})^2 + B \epsilon^{(m)}_{6} < 1.
\]

If we assume the error rates are the same for all types of locations ($\gamma=1$), we have $B=2,892$ and $A={196\choose 3}$, and Eq. (\ref{eq:error threshold bound}) gives
%\[
%{228\choose 3} (\epsilon^{(m)}_{6})^3 +6512 (\epsilon^{(m)}_{6})^2 < \epsilon^{(m)}_{6},
%\]
%\[
%{196\choose 3} (\epsilon^{(m)}_{6})^3 +2892 (\epsilon^{(m)}_{6})^2 < \epsilon^{(m)}_{6},
%\]
 an error threshold of
\[
\epsilon(\gamma=1)<3.06\times 10^{-4}.
\]
%For $\gamma<1$,
%Binomial[164, 3] + Binomial[164, 2]*Binomial[32, 1]*0.1 +
% Binomial[164, 1]*Binomial[32, 2]*0.01 + Binomial[32, 3]*0.001
%If we assume $\gamma=0.1$ or $ \gamma=0$,  the number of effective malignant pairs are $2185.9$ and $2118.0$,  and the error thresholds become
%\[\epsilon(\gamma=0.1)<4.06\times 10^{-4},\]
%and
%%If we assume no memory error,  the number of effective malignant pairs is $2118.0$ and the error threshold becomes
%\[\epsilon(\gamma=0)<4.14\times 10^{-4},\]
%respectively.
We compare our results with those of the Steane code and the Bacon-Shor code for $\gamma=1.0$ and $0.1$ in Table \ref{table:comparison concatenated codes}.
The rigorous error thresholds obtained in \cite{AGP06:QAT:2011665.2011666,AC:baconshor,quantum_accuracy_threshold} are also listed as a reference.
Knill's postselection scheme has the highest error threshold of $O(10^{-4})$, as expected.
Remark: we obtain 714 malignant pairs and calculate a threshold of $1.05\times 10^{-3}$ if we assume no SWAP or memory errors.
\begin{table}
\[
\begin{tabular}{|c|c|c|c|}
  \hline
 scheme & Steane code& Bacon-Shor code& Knill's postselection scheme\\
  \hline
nonlocal & $2.73\times 10^{-5}$ & $1.94\times 10^{-4}$ & $1.04\times 10^{-3}$ \\
2D($\gamma=1.0$) & $1.1\times 10^{-5}$  & $1.3\times 10^{-5}$ &  $3.06\times 10^{-4}$ \\
2D($\gamma=0.1$) & $1.85\times 10^{-5}$  & $2.02\times 10^{-5}$ &  $4.06\times 10^{-4}$ \\
  \hline
\end{tabular}
\]
\tcaption{Comparison of the error thresholds of three concatenated codes.} \label{table:comparison concatenated codes}
\end{table}

\subsection{Pseudo-Threshold}
Knill reported a simulated pseudo-threshold of $3\%$ by his postselection scheme over  unbiased and independent depolarizing noise~\cite{Knill05nature}.
We now present a Monte Carlo simulation of the circuit in Fig.~\ref{fig:1-exRec CNOT_ED+} over depolarizing noise to obtain pseudo-thresholds for the Knill scheme in two dimensions, as in~\cite{CDT09:CCS:2011814.2011815}.
In our model, we add depolarizing errors as quantum operations after gates or before measurements in the circuit.
Let $p$ be the depolarizing rate.  Any single qubit location (other than measurements) undergoes $X$, $Y$, or $Z$  with probability~$\frac{p}{3}$.
Any binary measurement outcome is flipped with probability $p$.
CNOT gates are modified by one of the $15$ non-identity two-qubit Pauli operators ($IX$, $IZ$, $\cdots$, $YY$) with probability~$\frac{p}{15}$.

We obtain the pseudo-thresholds by calculating the logical error rate of the circuit in Fig.~\ref{fig:1-exRec CNOT_ED+}.
The logical error rate $e(p)$ for a given depolarizing rate $p$ (the worst gate error rate)
is defined as the number of samples without logical errors at the output of the circuit in Fig.~\ref{fig:1-exRec CNOT_ED+},
divided by the number of samples without any errors being detected.
If an error-detecting code works, it is clear that $e(p)<p$ for $p$ small enough and $e(p)$ is an increasing function of $p$.
The pseudo-threshold $\tilde{\epsilon}$ is the value of $p$ such that $e(\tilde{\epsilon}^-)< \tilde{\epsilon}$ and $e(\tilde{\epsilon}^+)>\tilde{\epsilon}$.

If we assume all locations have the same depolarizing rate $p$ ($\gamma=1.0$), we find a pseudo-threshold of about $0.1\%$.
If $\gamma=0.1$,  we obtain a pseudo-threshold of about $0.2\%$.
These values are higher than the error thresholds estimated in the adversarial noise model, as expected, since the adversarial noise model is  the worst case.
As a comparison, we calculated a pseudo-threshold of about $0.8\%$ for the Knill scheme without locality constraints.

Remark: we can reduce the depolarizing rates on the measurements and ancilla preparations as Knill did in~\cite{Knill05nature}
by choosing these error rates to be $4/15$ of the worst gate error rate.
We obtain a pseudo-threshold of about $0.35\%$ by choosing $\gamma=0.1$ for the Knill scheme in two dimension.
For the Knill scheme without locality constraints, we obtain a pseudo-threshold of about $2.5\%$.

\section{Discussion}
\label{sec:conclusion}

We designed a two-dimensional $5\times 5$ qubit tile for quantum computation using the concatenated $C_4$ code with postselection.
Although we didn't prove the optimality of our design, we believe that a substantial improvement within our architectural framework is unlikely.

In this paper we demonstrated the tile operations of the ED$_{+}$ block for structure \textrm{I}.
Different combinations of error detection (ED$_{+}$ or ED$_{-}$) and  tile structures (\textrm{I} or \textrm{II})  %affect the
%The error detection ED$_+$ or ED$_0$ in a rectangle
 require small modifications to the logical gates involving two tiles, such as vSWAP, hSWAP, vCNOT, and hCNOT.
%since the ancilla preparations for ED$_+$ and ED$_0$ are different and they depend on the tile structures \textrm{I} or \textrm{II}.
These modifications  can be done by slightly changing the locations of $d_i, a_i, a_{i+4}$ for $i=1,2,3,4$ in our demonstration.
For example, we have to modify the ancilla preparation of an ED$_+$ that follows a vSWAP, and it takes one more time step  and four more lower-level SWAPs than a vSWAP followed by an ED$_0$.

%The memory error rate is very low in some physical structures, such as superconducting qubits.
%One would like to reduce the size of the tile and hence the number of SWAPs.
It is desirable to reduce the size of the tile, and hence the number of SWAPs, for physical architectures with very low memory
error rates, such as superconducting qubits.
To that end, we have also designed a $4\times 4$ tile, and its performance is compared with the $5\times 5$ tile in Table~\ref{table:tile_comparison}
for different ratios of memory error to gate error rate.
The $4 \times 4$ tile has a higher threshold with no memory error $(\gamma = 0)$.
Surprisingly, the error threshold of the $4 \times 4$ tile decreases by a factor of about two for $\gamma = 0.1$.
%Surprisingly, the error threshold of the $4\times 4$ tile for $\gamma=0$ increases to more than twice for $\gamma=0.1$.
%is higher than the $4\times 4$ tile even w.
This is probably because there are many more idle qubits in the $4\times 4$ tile, and the operations in the two code blocks of the $4\times 4$ tile are not parallel: one block is delayed by one time step as shown in the tile operations online.
However, the error thresholds of the $5\times 5$ tile for $\gamma=0.1$ and $\gamma=0$ are about the same.
The effects of some errors may cancel each other due to the symmetry in the 1-exRec of the CNOT gate in the $5\times 5$ tile.

%%\[
%%\alpha=\begin{pmatrix}
%%34&	39&	42&	46&	84&	104&	138\\
%%&	&	39&	55&	103&	126&	175\\
%%&	&	12&	42&	76&	124&	146\\
%%&	&	&	24&	96&	157&	199\\
%%&	&	&	&	98&	738&	364\\
%%&	&	&	&	&	533&	1526\\
%%&	&	&	&	&	&	375\\
%%\end{pmatrix}.
%%\]
%%The threshold is $1.71544\times 10^{-4}$.
%%If the ration is 0.1/0.0, mal eff is 2.830550000000000e+03/2572.
%%The threshold is $3.29779\times 10^{-4}$
%%$3.58907\times 10^{-4}$

%\begin{table*}
%\[
%\begin{tabular}{|c|c|c|c|c|c|c|}
%  \hline
%  % after \\: \hline or \cline{col1-col2} \cline{col3-col4} ...
%&\multicolumn{3}{c}{1-exRec of the CNOT gate}\vline&\multicolumn{3}{c}{effective malignant pairs and error threshold} \vline\\
%\hline
%  tile & SWAPs & idle qubits& times steps & $\gamma=1$& $\gamma=0.1$& $\gamma=0.0$ \\
%  \hline
%% $4\times 4$ & 38  & 74  & 16  & 5495~\vline~$1.72\times 10^{-4}$ & 2830.6~\vline~$3.30\times 10^{-4}$  & 2572~\vline~$3.59\times 10^{-4}$ \\
%%$5 \times 5$ & 48  & 32  & 14  & 4032~\vline~$2.22\times 10^{-4}$ & 2151.0~\vline~$4.06\times 10^{-4}$  & 1968~\vline~$4.38\times 10^{-4}$\\
% $4\times 4$ & 38  & 74  & 16  & 4148~\vline~$1.47\times 10^{-4}$ & 1572.1~\vline~$2.22\times 10^{-4}$  & 1443~\vline~$4.89\times 10^{-4}$ \\
%$5 \times 5$ & 48  & 32  & 14  & 2892~\vline~$3.06\times 10^{-4}$ & 2185.9~\vline~$4.06\times 10^{-4}$  & 2118~\vline~$4.14\times 10^{-4}$\\
%  \hline
%\end{tabular}
%\]
%\fcaption{Comparison of the $4\times 4$ and $5\times 5$ tiles.} \label{table:tile_comparison}
%\end{table*}
\begin{table*}
\[
\begin{tabular}{|c|c|c|c|c|c|c|}
  \hline
  % after \\: \hline or \cline{col1-col2} \cline{col3-col4} ...
&\multicolumn{3}{c}{1-exRec of the CNOT gate}\vline&\multicolumn{3}{c}{error threshold} \vline\\
\hline
  tile & SWAPs & idle qubits& times steps & $\gamma=1$& $\gamma=0.1$& $\gamma=0.0$ \\
  \hline
% $4\times 4$ & 38  & 74  & 16  & 5495~\vline~$1.72\times 10^{-4}$ & 2830.6~\vline~$3.30\times 10^{-4}$  & 2572~\vline~$3.59\times 10^{-4}$ \\
%$5 \times 5$ & 48  & 32  & 14  & 4032~\vline~$2.22\times 10^{-4}$ & 2151.0~\vline~$4.06\times 10^{-4}$  & 1968~\vline~$4.38\times 10^{-4}$\\
 $4\times 4$ & 38  & 74  & 16  & $1.47\times 10^{-4}$ & $2.22\times 10^{-4}$  & $4.89\times 10^{-4}$ \\
$5 \times 5$ & 48  & 32  & 14  & $3.06\times 10^{-4}$ & $4.06\times 10^{-4}$  & $4.14\times 10^{-4}$\\
  \hline
\end{tabular}
\]
\tcaption{Comparison of the $4\times 4$ and $5\times 5$ tiles.} \label{table:tile_comparison}
\end{table*}

Under the realistic assumption that one- and two-qubit quantum gates are local,
our threshold analyses establish that Knill's postselection
scheme has better error correction capabilities than other concatenated error-correcting codes.
This makes our proposed two-dimensional architecture a practical choice for quantum error correction.

In addition to the postselection scheme based on error detection, Knill also proposed a Fibonacci scheme to further reduce the overhead of the postselection scheme~\cite{Knill05nature}.
He calculated a pseudo-threshold of about $1\%$.
It uses the fact that the concatenated error-detecting code $C_4$ can correct located errors.
Aliferis and Preskill showed that the error threshold of the Fibonacci scheme is slightly lower than the postselection scheme over the adversarial noise model~\cite{AP09}.
Nonrecursive versions of the CNOT gates or the measurements in the Fibonacci scheme
would take many time steps without error detection or correction in our two-dimensional architecture.
This might lead to a much worse error threshold.
However, we still consider finding the threshold of the Fibonacci scheme, combined with the ``soft decision" decoder in \cite{ES12}, an interesting question for future work.
%This is our future work.

%We refer interested readers to our technical report \cite{QCS_tech12} for resources estimation of the physical implementation of some quantum algorithms on this two-dimensional architecture.

\nonumsection{Acknowledgements}
\noindent
This work was supported by the  Intelligence Advanced Research Projects Activity (IARPA) via Department of Interior National Business Center contract numbers D11PC20165 and D11PC20167.
The U.S. Government is authorized to reproduce and distribute reprints for Governmental purposes notwithstanding any copyright annotation thereon.
The views and conclusions contained herein are those of the authors and should not be interpreted as necessarily representing the official policies or endorsements, either expressed or implied, of IARPA, DoI/NBC, or the U.S. Government.
%This research was supported by the Intelligence Advanced Research Projects Activity (IARPA) via Department of Interior National Business Center contract number D11PC20165.
%The U.S. Government is authorized to reproduce and distribute reprints for Governmental purposes notwithstanding any copyright annotation thereon.
%The views and conclusions contained herein are those of the authors and should not be interpreted as necessarily representing the official policies or endorsements, either expressed or implied, of IARPA, DoI/NBC, or the U.S. Government.

\nonumsection{References}
%\noindent
%{\small
%\bibliographystyle{IEEEtran}
%\bibliography{RefEstimates}
%}

% Generated by IEEEtran.bst, version: 1.13 (2008/09/30)

%\begin{thebibliography}{000}
%\bibitem{cal}
%R. Calderbank and P. Shor (1996), {\it Good quantum error
%       correcting codes exist},
%Phys. Rev. A, 54, pp. 1098-1106.
%
%\bibitem{niel}
%M.A. Nielsen and J. Kempe (2001), {\it Separable states are
%more disordered globally than locally}, quant-ph/0105090.
%
%\bibitem{mar}
%A.W. Marshall and I. Olkin (1979), {\it Inequalities: theory of majorization and its applications},
%Academic Press (New York).
%\end{thebibliography}

\end{document}